\documentclass[prl,aps,reprint,showpacs,groupedaddress,superscriptaddress,twocolumn,final,prabib]{revtex4-1}
\usepackage{graphicx}
\usepackage{amsmath}
\usepackage{amssymb}
\usepackage{verbatim}
\usepackage{hyphenat}
\usepackage{epstopdf}

\begin{document}

\title{Disorder induced power-law response of a superconducting vortex on a plane}

\author{N. Shapira}
\affiliation{Department of Physics, Technion - Israel Institute of Technology, Haifa, 32000, Israel}
\author{Y. Lamhot}
\affiliation{Department of Physics, Technion - Israel Institute of Technology, Haifa, 32000, Israel}
\author{O. Shpielberg}
\affiliation{Department of Physics, Technion - Israel Institute of Technology, Haifa, 32000, Israel}
\author{Y. Kafri}
\affiliation{Department of Physics, Technion - Israel Institute of Technology, Haifa, 32000, Israel}
\author{B. J. Ramshaw}
\affiliation{Department of Physics and Astronomy, University of British Columbia, Vancouver, British Columbia V6T 1Z1, Canada}
\author{D. A. Bonn}
\affiliation{Department of Physics and Astronomy, University of British Columbia, Vancouver, British Columbia V6T 1Z1, Canada}
\author{Ruixing Liang}
\affiliation{Department of Physics and Astronomy, University of British Columbia, Vancouver, British Columbia V6T 1Z1, Canada}
\author{W. N. Hardy}
\affiliation{Department of Physics and Astronomy, University of British Columbia, Vancouver, British Columbia V6T 1Z1, Canada}
\author{O. M. Auslaender}
\affiliation{Department of Physics, Technion - Israel Institute of Technology, Haifa, 32000, Israel}

\date{\today }

\begin{abstract}
We report drive-response experiments on individual superconducting vortices on a plane, a realization for a 1+1-dimensional directed polymer in random media. For this we use magnetic force microscopy (MFM) to image and manipulate individual vortices trapped on a twin boundary in YBCO near optimal doping. We find that when we drag a vortex with the magnetic tip it moves in a series of jumps. As theory suggests the jump-size distribution does not depend on the applied force and is consistent with power-law behavior. The measured power is much larger than widely accepted theoretical calculations.
\end{abstract}

\maketitle


While the dynamics of driven systems in ordered media are well understood, disorder gives rise to much more elaborate behavior. Particularly interesting are phenomena arising from the interplay between disorder and elasticity \cite{agoritsas2012disordered,kolton2006dynamics} such as the conformations of polyelectrolytes \cite{de1979scaling} (e.g. polypeptides and DNA \cite{bustamante2003ten}), kinetic roughening of driven interfaces (e.g. wetting in paper \cite{halpin1995kinetic,herminghaus2012universal}, magnetic and ferroelectric domain wall motion \cite{lemerle1998domain,paruch2005domain,kim2009interdimensional,yamanouchi2007universality}, the growth of bacterial colony edges \cite{halpin1995kinetic}), non-equilibrium effects that occur in randomly stirred fluids \cite{forster1977large} and more. Superconducting vortices, in materials in which they behave like elastic strings, are among the most important examples of such systems \cite{blatter1994vortices}. Despite a dearth of direct experimental proof, these quantized whirlpools of supercurrent are considered textbook examples of the theory of directed polymers in random media (DPRM) \cite{kardar2007statistical,halpin20122+,dotsenko2008joint}, a foundation model for systems where disorder and elasticity compete. This model, that yields many results that are considered generic and universal, provides the backdrop for our experiment.

\begin{figure*}
\centering
\setlength\fboxsep{0pt}
\setlength\fboxrule{0.25pt}
\includegraphics[width=7in]{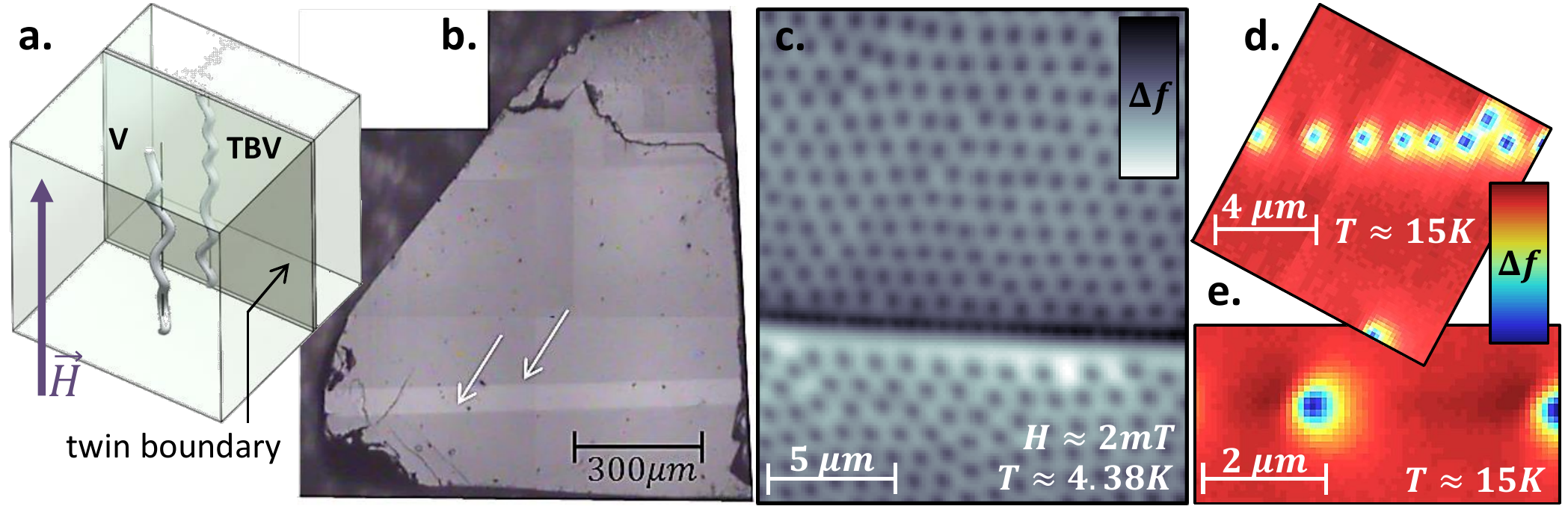}
\caption{Vortices on and off a twin boundary (TB). {\bf(a)} Illustration of vortices in a superconductor. The left vortex (V) can meander in $d_\perp=2$ dimensions perpendicular to an external magnetic field $\vec{H}$  while the vortex trapped on the TB (TBV), a common planar defect, can meander only on the plane, i.e. in $d_\perp=1$ dimensions. {\bf(b)} Polarized-light microscopy photo of our $80\mu m$-thick sample revealing two TBs (white arrows). {\bf(c)} MFM scan of vortices (black discs) that form a high density chain along a TB and an Abrikosov lattice around it ($z\approx1.15\mu m$, $\Delta f$ spans $0.93Hz$). {\bf(d)} MFM scan at $0\leq H\leq10\mu T$. Vortices (blue discs) accumulate on a TB and exhibit $1+1$-dimensional physics ($z\approx0.28\mu m$, $\Delta f$ spans $1.6Hz$). {\bf(e)} Many vortices in the chain in (d) are isolated because their separation is much larger than $\lambda_{ab}\approx120nm$ (here $z\approx0.4\mu m$, $\Delta f$ spans $0.6Hz$).}
\label{fig:fig1}
\end{figure*}

Here we concentrate on vortices that are trapped on a twin boundary (TB), a planar defect in YBa$_2$Cu$_3$O$_{7-\delta}$ (YBCO) \cite{nam2005twinning}. We cool the sample through the superconducting transition temperature $T_c$ in the presence of an external magnetic field $\vec{H}=H\hat{z}$, which directs the curve along which vortices cross the sample. Figure~\ref{fig:fig1}a depicts a vortex away from a TB (V in Fig.~\ref{fig:fig1}a) that is free to meander in the $d_{\perp}=2$ directions perpendicular to $\vec{H}$. For a vortex trapped on a TB (TBV in Fig.~\ref{fig:fig1}a) the meandering is limited to a plane, i.e. $d_{\perp}=1$.  We concentrate on TB-vortices both because the reduced dimensionality makes data analysis simpler and, more importantly, because, unlike DPRM in higher dimensions, $1+1$-DPRM is a tractable model \cite{hwa1999mesoscopic}.

The path of a vortex across a sample is determined by the competition between elasticity and disorder: while meandering allows a vortex to lower the energy of the system by locating its core near defects, the associated stretching is limited by finite line tension $\kappa$ \cite{blatter1994vortices}. As a result the unavoidable random disorder in a sample can make the optimal path for an isolated vortex elaborate. Despite this, DPRM theory provides many predictions for disorder-averaged quantities \cite{hwa1994anomalous}. For example, the thermal and disorder averaged offset distance from the field axis $\hat z$ ($\Delta$) scales like a power-law given by the wandering exponent $\zeta(d_\perp)$: $\overline{\left<\Delta\right>}\equiv\delta R\sim L^{\zeta(d_\perp)}$ for $L\gg a_z$ ($L$ is the sample thickness, $a_z$ is a sample-dependent lower-cutoff), which is a universal number. Theoretically $\zeta(d_\perp)$ describes a wide variety of systems \cite{kardar2007statistical} but there are only a few measurements of it \cite{lemerle1998domain,paruch2005domain,kim2009interdimensional,yamanouchi2007universality,bolle1999observation,takeuchi2010universal,takeuchi2011growing}.
While a power-law also describes classical random walks ($\delta R\sim L^{\frac{1}{2}}$) disorder both enhances wandering ($\zeta(d_\perp)>\frac{1}{2}$) and stretches the distribution of offset distances $W(\Delta)$ from gaussian to $W(\Delta)\sim\Delta^{-\alpha_{theory}}$ ($\alpha_{theory}>0$), significantly increasing the prevalence of trajectories with large excursions \cite{hwa1994anomalous}.

The power-law form of $W(\Delta)$ implies the absence of a characteristic length-scale and the existence of a significant number of vortex trajectories with a wide variety of $\Delta$'s and with free-energies almost as low as that of the optimal vortex path. Since these trajectories constitute the low-energy excitations of the system they are important for thermodynamics and response functions \cite{hwa1994anomalous}. While in thermal equilibrium the system has time to find these metastable states it is not clear what happens out of equilibrium, although one can expect that near equilibrium these trajectories remain important.

In this work we experimentally characterize the trajectories of individual vortices confined to move on a TB. Unlike most previous work we use a local probe (magnetic force microscopy, MFM) to measure individual vortices. The heart of MFM is a sharp magnetic tip situated at the end of a cantilever driven to oscillate in the $z$-direction normal to the sample surface at a resonant frequency $f$. A force $\vec{F}=F_x {\hat x}+F_y{\hat y}+F_z{\hat z}$ acting on the tip shifts $f$ by $\Delta f=f-f_0\approx-f_0/(2k)\partial F_z/\partial z$ ($f_0$ is the natural resonant frequency, $k$ is the cantilever spring constant, $z$ is the tip-sample distance) \cite{albrecht1991frequency}. For an image we record $\Delta f$ while rastering the tip at constant $z$. In addition we use the tip-vortex interaction to perturb vortices individually \cite{straver2008controlled}. Such perturbations show up as abrupt shifts of the signal from a vortex, which we dub "jumps".

The sample we used is a nearly optimally doped YBCO single crystal ($T_c\approx91K$ \cite{TcNote}) grown from flux in a BaZrO$_3$ crucible for high purity and crystallinity \cite{liang1998growth}. The $L=80\mu m$-thick platelet-shaped sample has faces parallel to the crystal ab-plane and contains two parallel TBs (Fig.~\ref{fig:fig1}b). Field cooling was done with $\vec{H}=H\hat{z}$ parallel to the crystal $c$-axis and along the TB plane with the tip magnetized for attractive tip-vortex interactions.

Figure~\ref{fig:fig1}c is an MFM scan of vortex arrays on a TB and around it for $H\approx2mT$. Such a highly ordered Abrikosov lattice \cite{abrikosov1957magnetic,kleiner1964bulk} at such a low field attests to the scarcity of strong defects other than the TB. Figure~\ref{fig:fig1}d is an MFM scan of a TB at $0\leq H\leq10\mu T$. In this near-zero field almost all of the vortices were trapped by the TBs, further attesting to the high quality of the sample and in agreement with early experiments showing that in YBCO TBs are strong vortex traps \cite{vinnikov1988direct}. Despite their relative high density, many of the TB-vortices can be considered isolated since their nearest-neighbor distance is much larger than the penetration depth $\lambda_{ab}\approx120nm$ \cite{kiefl2010direct} (Fig.~\ref{fig:fig1}e).

We tested how strongly vortices are trapped by a TB by performing low-height (and hence strong lateral force, up to 20 pN) scans. However, even for our lowest passes across the TB and even for $T\approx0.85T_c$ we never observed a vortex dislodging from the TB. This experimentally verifies that for the range of forces we applied TB-vortices behave as one-dimensional (1D) objects in an effective $d=1+1$ geometry.

Next, we performed a series of raster scans over an isolated TB-vortex (Fig.~\ref{fig:fig1}e) in order to perturb it. The scan pattern consisted of line-scans in which the tip moved back and forth (Fwd/Bwd) at $125\frac{nm}{sec}$ along the $x$-axis parallel to the TB. After each line-scan we reduced $z$ and stepped the tip in the $y$-direction. Since the force the tip exerts on a vortex depends on both $z$ and the tip-vortex lateral distance $\rho=\sqrt{(x-x_v)^2+(y-y_v)^2}$ ($x_v{\hat x}+y_v{\hat y}$ is the vortex position in the scan, see \cite{ForceNote}), a complete scan series gives the response of a TB-vortex to a wide range of forces along the TB, $F_x$.

\begin{figure}[h]
\includegraphics[width=3.4in]{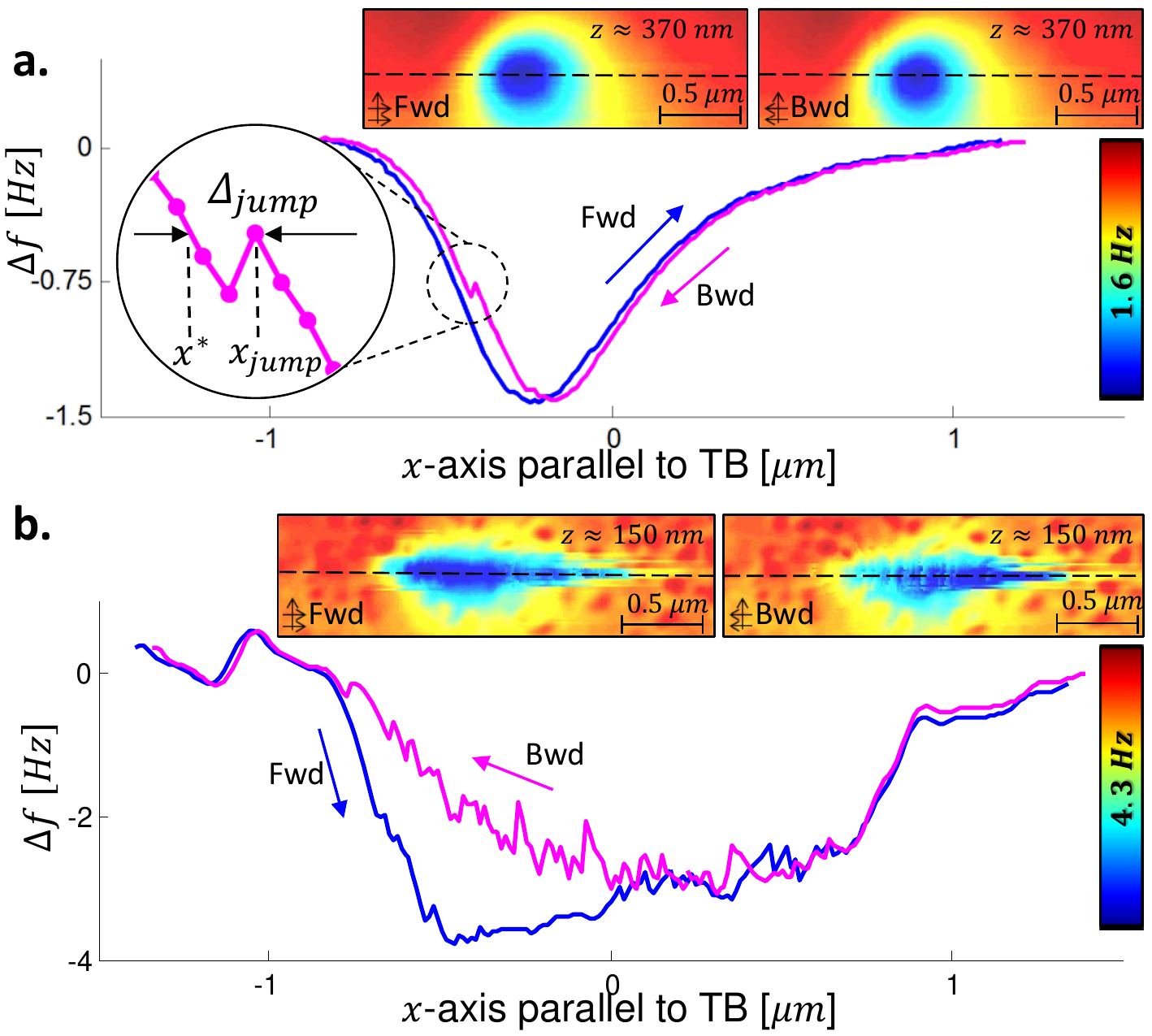}
\caption{Manipulation scans of TB-vortices at $T=15K$. {\bf(a)} Forward (Fwd) and backward (Bwd) line-scans (taken along the dashed lines from the scans in the insets) containing a tip-induced vortex jump of size $\Delta_{jump}=|x^*-x_{jump}|$ that we associate with an abrupt change in the position of the upper part of the vortex. {\bf(b)} Fwd and Bwd line-scans taken along the dashed lines from the scans in the insets. Numerous vortex jumps with a variety of $\Delta_{jump}$'s are apparent. The difference between the overall shapes of the Fwd and Bwd line-scans suggests that non-equilibrium effects may be involved. {\bf Insets:} The scans from which the line-scans in (a) and (b) were taken. The scan height and the span of $\Delta f$ is indicated for each panel. The horizontal double arrows indicate the back or forth scan direction along the TB (the x-axis) and the large vertical arrows indicate the direction we step the tip after each back and forth cycle (the y-axis).}
\label{fig:fig2}
\end{figure}

Figure~\ref{fig:fig2}a shows typical line-scans for an almost static vortex. $\Delta f$ becomes increasingly negative as the tip approaches the vortex due to the increasing tip-vortex attraction until it passes the minimal $\rho$ in the line-scan. From there $\Delta f $ increases until the interaction becomes negligible again. The line-scans in Fig.~\ref{fig:fig2}a show one of the first jumps for this particular vortex - a shift in $\Delta f(x)$  at $x=x_{jump}$. We associate this shift with a tip-induced abrupt change in the position of the upper part of the vortex. We determine the jump length $\Delta_{jump}=|x_{jump}-x^*|$  from the first position after the jump satisfying $\Delta f(x^*)=\Delta f(x_{jump})$ \cite{AlgorithmNote}. In addition, we calculate the value of $F_x$ before each jump using an approximation for the magnetic field from a single vortex and a model for the tip \cite{ForceNote}. Figure~\ref{fig:fig2}b shows typical line-scans for a moving vortex. While the signal in the central region of the line-scan contains numerous sharp changes, the envelope resembles a stretched version of the signal expected from a static vortex at the same $z$. This indicates that in the central region the top of the vortex moves with the tip in a series of jumps. The observed asymmetry between the Fwd and Bwd line-scans are typical for a moving vortex and suggest that the system is not in thermal equilibrium.

\begin{figure}[h]
\includegraphics[width=2.8in]{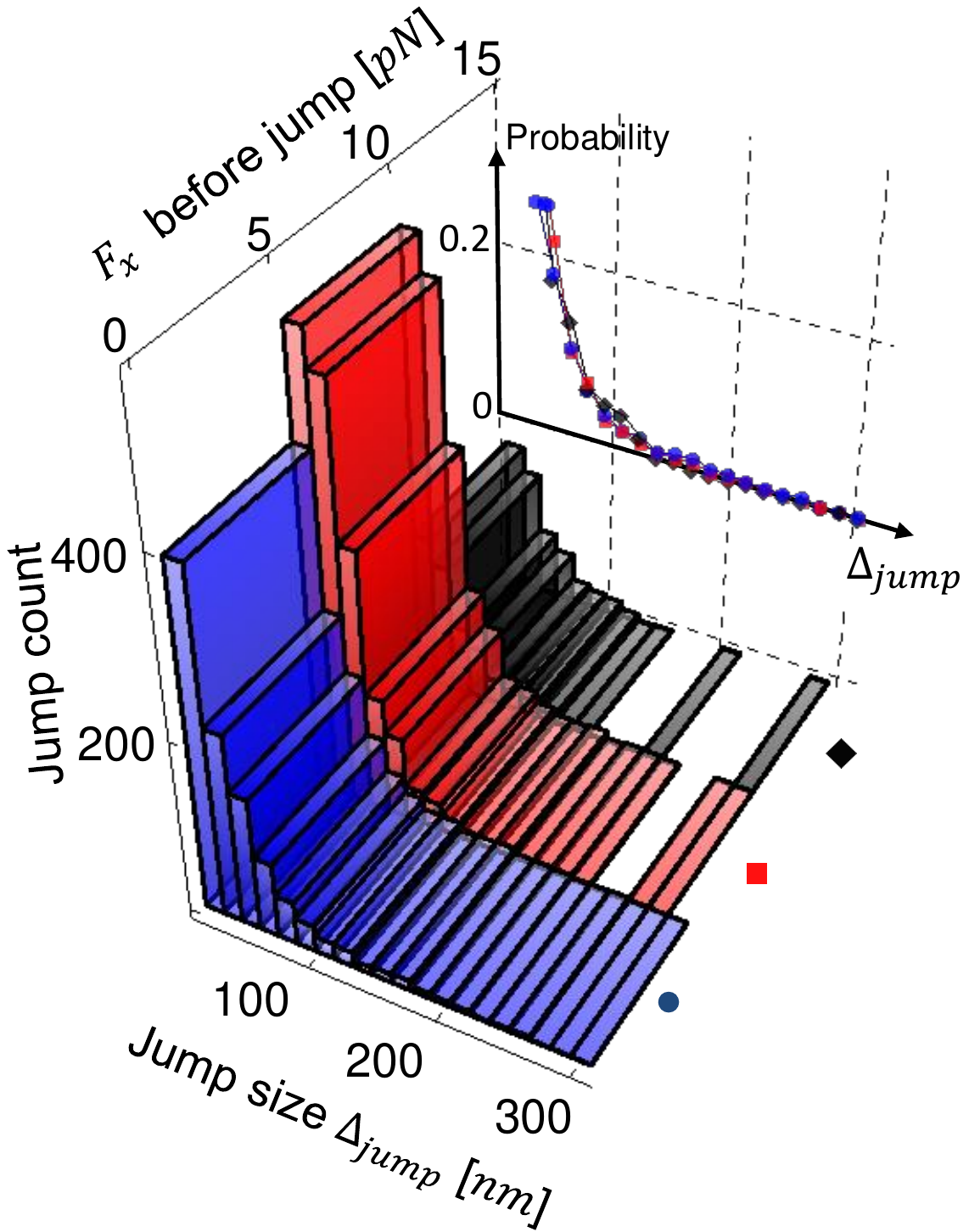}
\caption{Histograms binning all measured jump lengths ($\Delta_{jump}$) for different ranges of the force exerted along the TB by the tip ($F_x$). \textbf{Inset:} Normalized distributions of $\Delta_{jump}$ for the different $F_x$ ranges. All the distributions collapse onto each other revealing the independence of $\Delta_{jump}$ from $F_x$.}
\label{fig:fig3}
\end{figure}

Figure~\ref{fig:fig3} shows histograms containing all jumps of two vortices chosen for their large separation from their neighbors and each other (enough to safely consider their disorder environments independent). The histograms separate the jumps into three ranges of $F_x$. When we compare the distribution of $\Delta_{jump}$ within each $F_x$ range we find that the distributions collapse onto each other. Moreover, we find the same collapse when we consider jumps from each vortex separately \cite{SelfAVGNote}. This shows that for the range of forces we applied the distribution of $\Delta_{jump}$ does not depend on $F_x$ and justifies lumping all the jumps together regardless of the force.

\begin{figure}[h!]
\includegraphics[width=3.4in]{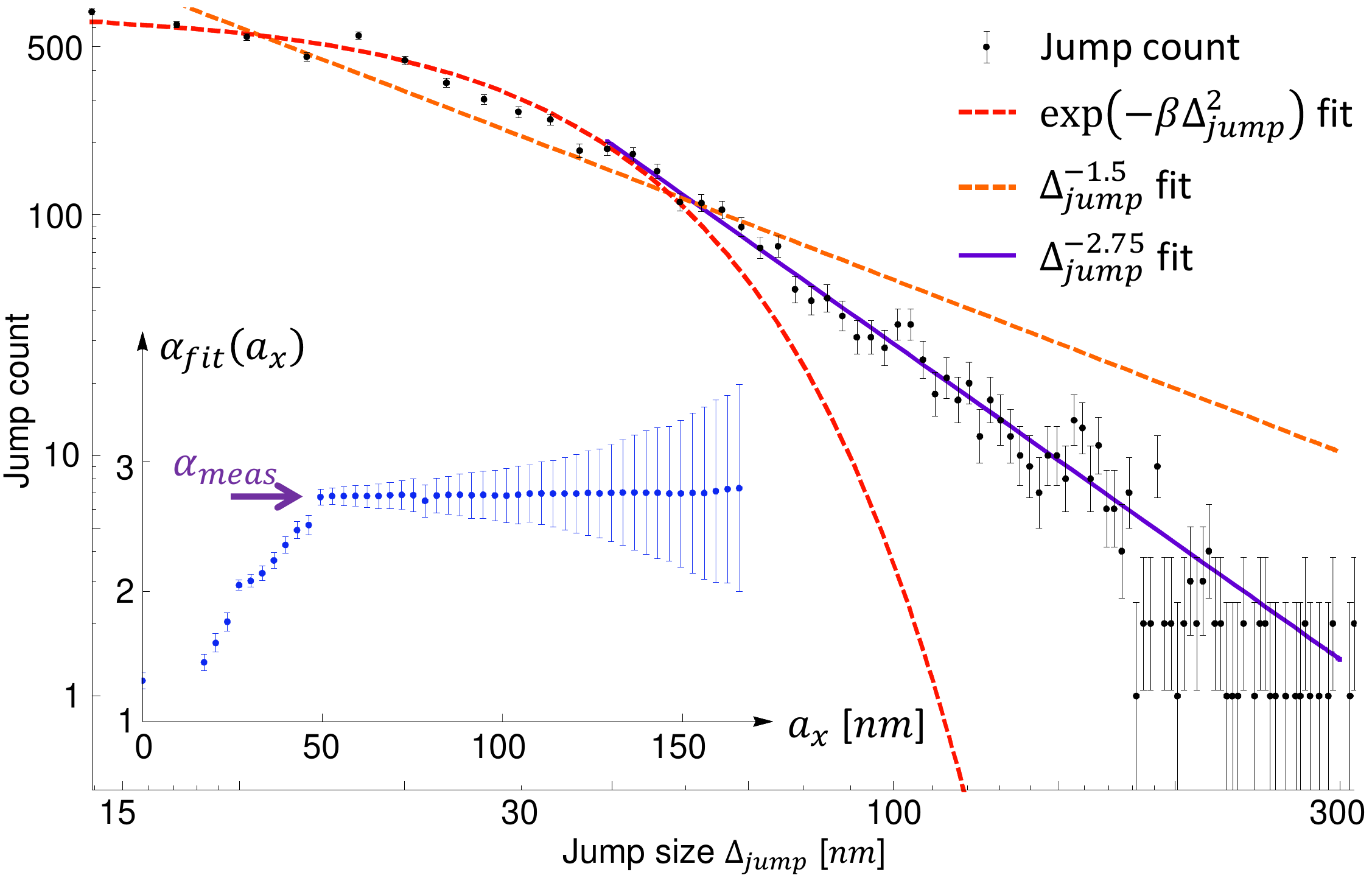}
\caption{Measured vortex jump lengths ($\Delta_{jump}$) and fits to the data. Although the data is consistent with a power-law distribution the exponent we obtain does not match $\alpha_{theory}=3/2$ predicted for a system in equilibrium. \textbf{Inset:} Values of a power-law exponent $\alpha_{fit}$ obtained by fitting the data in the main panel for different values of the lower cutoff $a_x$. $\alpha_{fit}$ saturates (arrow) for $a_x>a_0=49\pm3$, a clear indication that $\alpha_{meas}=2.75\pm0.06$ is the best-fit exponent for the distribution.}
\label{fig:fig4}
\end{figure}

Our main result is the force-independent distribution $\tilde{W}(\Delta_{jump})$ for both vortices together (Fig.~\ref{fig:fig4}). The most significant feature of $\tilde{W}(\Delta_{jump})$ is a long tail indicating that disorder is important - it is in complete disagreement with the gaussian distribution expected for a system where disorder is irrelevant \cite{hwa1994anomalous}. Another important feature is the saturation of $\alpha_{fit}$ obtained from best fits of $\tilde{W}(\Delta_{jump})$ to a power-law for different values of a lower cutoff $a_x$. The saturation is a strong indication that $\tilde{W}(\Delta_{jump})$ is a power-law for $\Delta_{jump}>a_0=49\pm3nm$ with the power given by $\alpha_{meas}=2.75\pm0.06$ ($80\%$ confidence level). We emphasize that we determined $\Delta_{jump}$ directly and without theoretical assumptions  and that $\tilde{W}(\Delta_{jump})$, $\alpha_{meas}$ and $a_0$ are not sensitive to several important sources of systematic error (the independence of $\tilde{W}(\Delta_{jump})$ on $F_x$ implies that it is not sensitive to systematic errors in force estimates, the scale invariance of power-laws implies that $\alpha_{meas}$ is insensitive to errors in length calibration).

According to the fluctuation-susceptibility relation \cite{hwa1994anomalous} the statistics of the jump length ($\Delta_{jump}$) gives information on the properties of the rare, large-scale, low-energy excitations of the system characterized by $\Delta$. One might worry that when driven out of equilibrium short jumps will occur more readily than the long jumps required to reach one of the more favorable paths farther away. However, the properties of the accessible vortex trajectories ensure that such behavior is unlikely \cite{hwa1994anomalous}.

The independence of $\tilde{W}(\Delta_{jump})$ on $F_x$ (Fig.~\ref{fig:fig3}), which at first glance may seem puzzling, is attributed  by DPRM theory \cite{hwa1994anomalous} to \textit{statistical tilt symmetry}. This symmetry is a manifestation of the absence of correlations in the disorder which means that for sufficiently large force \cite{hwa1994anomalous}, as in this experiment \cite{MinForceNote}, each time we tilt a vortex it samples a new random environment and is equivalent to an un-tilted vortex experiencing a new disorder realization. The observed statistical tilt symmetry implies that theoretically we could have obtained disorder-averaged quantities from measurements of just one vortex. Indeed, when we examine the force-independent distributions of $\Delta_{jump}$ for each vortex separately \cite{TwoVortexNote} we find that the distributions are statistically similar. This observed self-averaging corroborates the statistical tilt symmetry and means that the measured distribution of jump lengths is indeed equivalent to the distribution of rare, large-scale, low-energy excitations; i.e. $\tilde{W}(\Delta_{jump})=W(\Delta)$.

While DPRM predicts the power-law behavior of $W(\Delta)$, the value we extract disagrees with the theoretical value: $\alpha_{theory}=d_{\perp}+2-\zeta^{-1}(d_\perp)$ \cite{hwa1994anomalous}. The value of the wandering exponent $\zeta(d_\perp=1)=2/3$ has been theoretically found by various methods \cite{kardar1987replica,huse1985huse,gwa1992bethe} giving $\alpha_{theory}=3/2$, very different from $\alpha_{meas}\approx2.75$. This deviation could result from a variety of reasons; however, the asymmetry of the line traces in Fig.~\ref{fig:fig2}b suggests that non-equilibrium effects may be involved. The fact that we observe a response that remains power-law distributed even out of equilibrium is surprising. Whether or not non-equilibrium effects in fact explain the enhancement of $\alpha_{meas}$ is a question that requires further study.

The value of the cutoff $a_0$ provides a new way to characterize statistical properties of the disorder near a TB. This is due to general scaling arguments that hold both in and out of equilibrium \cite{hwa1994anomalous} and give a relationship between $a_0$ and the disorder strength $D$, that in $d=1+1$ is $D=(k_BT)^3/(a_0\kappa)$ ($k_B$ is the Boltzmann constant) \cite{hwa1994anomalous}. Using $T=15K$ and $\kappa=2.4eV/\mu m$ we find $\sqrt{D}\approx135\mu eV$ \cite{KappaNote}. Similar scaling relations give an estimate for the cutoff along $z$, i.e. $a_z=(a_0^2\kappa)/(k_BT)\approx4.5\mu m\ll L=80\mu m$, consistent with the experiment being in the thick sample regime.

To conclude, we have used the interaction between a magnetic tip and superconducting vortices on a TB to study the behavior of individual directed 1D objects. This provides an ideal setup for studying the interplay between elasticity and disorder, which is ubiquitous in nature. After experimentally showing that vortices on a TB behave as 1D objects in an effective $1+1$ random medium we proceeded to pull them one at a time along the TB and measured the distribution of jump lengths $\tilde{W}(\Delta_{jump})$. We find that $\tilde{W}(\Delta_{jump})$ is independent of the force applied by the tip and is the same for two widely separated vortices, confirming the predicted statistical tilt symmetry in the system. Our central result is the power-law form of $\tilde{W}(\Delta_{jump})$ that suggests that even out of equilibrium excitations do not have a characteristic length-scale beyond the sample-specific lower cutoff $a_0$. The direct measurement of $a_0$ provides a new characterization of the local disorder strength $D$ around the TB, complementing other measures such as the critical current \cite{larbalestier2001high,wee2013engineering}.

We thank Thierry Giamarchi, who encouraged us to focus on vortex motion along the TB,  as well as Anatoli Polkovnikov and Daniel Podolsky for comments and Gad Koren for help with characterization. N.S. acknowledges support from the Gutwirth Fellowship and Posnansky Research Fund in High Temperature Superconductivity. O.M.A. is supported by an Alon Fellowship and as a Horev Fellow is supported by the Taub Foundation. The project has received funding from the European Community’s Seventh Framework Programme (FP7/2007-2013) under Grant Agreement n$^\circ$ 268294.

\bibliography{PowerLaw}

\end{document}